\documentclass[12pt]{article}
\usepackage{a4wide}
\usepackage{bbm}
\usepackage{amsmath}
\usepackage{amssymb}

\newcommand\Frac[2]{{\textstyle\frac{#1}{#2}}}
\newcommand{\1}{\mathbbm{1}}
\newcommand{\C}{\mathbbm{C}}
\newcommand{\E}{\mathbbm{E}}
\renewcommand{\H}{\mathbbm{H}}
\newcommand{\N}{\mathbbm{N}}
\renewcommand{\P}{\mathbbm{P}}
\newcommand{\R}{\mathbbm{R}}
\newcommand{\Z}{\mathbbm{Z}}
\newcommand{\Lor}{\mathop{\rm Lor}\nolimits}
\newcommand{\lOR}{\mathop{\rm lor}\nolimits}
\newcommand{\Bor}{\mathop{\rm Bor}\nolimits}
\newcommand{\bor}{\mathop{\rm bor}\nolimits}
\newcommand{\Tor}{\mathop{\rm Tor}\nolimits}
\newcommand{\tor}{\mathop{\rm tor}\nolimits}
\newcommand{\Tra}{\mathop{\cal T}\nolimits}
\newcommand{\Mor}{\mathop{\rm Mor}\nolimits}
\newcommand{\rad}{\mathop{\rm rad}\nolimits}
\newcommand{\Rad}{\mathop{\rm Rad}\nolimits}
\newcommand{\Der}{\mathop{\rm Der}\nolimits}
\newcommand{\id}{\mathop{\rm id}\nolimits}
\newcommand{\rank}{\mathop{\rm rank}\nolimits}
\newcommand{\GL}{\mathop{\rm GL}\nolimits}
\newcommand{\SL}{\mathop{\rm SL}\nolimits}
\newcommand{\sL}{\mathop{\rm sl}\nolimits}
\newcommand{\SO}{\mathop{\rm SO}\nolimits}
\newcommand{\sO}{\mathop{\rm so}\nolimits}
\newcommand{\su}{\mathop{\rm su}\nolimits}
\newcommand{\Spec}{\mathop{\rm Spec}\nolimits}
\newcommand{\tr}{\mathop{\rm tr}\nolimits}
\newcommand{\diag}{\mathop{\rm diag}\nolimits}
\newcommand{\eps}{\varepsilon}
\newcommand{\rot}{\mathop{\rm rot}\nolimits}
\newcommand{\kest}{\mathop{\rm span}\nolimits}
\newcommand{\sol}{\mathop{\rm sol}\nolimits}
\newcommand{\po}{\raise7pt\hbox to0pt{$\scriptstyle\,\circ$\hss}p}

\begin{document}

\thispagestyle{empty}

\begin{center}
{\Large\bf Zero mass as the Borel structure}\\[1.0cm]
{\large Rein Saar and Stefan Groote}\\[0.3cm]
  Loodus- ja t\"appisteaduste valdkond, F\"u\"usika Instituut,\\[.2cm]
  Tartu \"Ulikool, W.~Ostwaldi 1, 50411 Tartu, Estonia
\end{center}

\vspace{0.2cm}
\begin{abstract}\noindent
The Lorentz group $\Lor_{1,3}=\SO_0(1,3)$ has two point fixgroups, namely
$\SO(3)$ for time-like translations and $\SO_0(1,1)\times\R^2$ for light-like
translations. However, for light-like translations it is reasonable to
consider a line fixgroup that leads to the Borel structure of the Lorentz
group and gives appropriate helicities for massless particles. Therefore,
whether a particle is massless or massive is not so much a physical question
but rather a question of the underlying Lie group symmetry.
\end{abstract}

\vspace{12pt}\noindent

\vspace{12pt}\noindent

\newpage

\section{Introduction\label{sec1}}
The basic structure of physics is defined via the representations of Lie
groups and Lie algebras. Symmetry arises via both the external or
space-like and internal or charge-like degrees of freedom and is represented
in complex vector spaces regarding the noncompact--compact dichotomy of the
relevant groups. The groups enter in two different varieties, the semisimple
and the solvable, as the building blocks for all other Lie groups. Simple Lie
groups regenerate themselves under commutation and generate semisimple groups
via direct products. Solvable Lie groups do not regenerate themselves under
commutation, but are constructed in a stepwise way out of Abelian subgroups.
Non-semisimple Lie groups are semidirect products of semisimple and solvable
subgroups. Note that there is no complete classification for solvable Lie
groups and therefore for non-semisimple Lie groups.

The subgroup structure of the symmetry group plays a basic role for Wigner's
definition of particles in the electroweak standard model, e.g.\ the
isotropy subgroups of the Lorentz group and their corresponding coset
manifolds. In particular, for massive particles the point fixgroup $\SO(3)$
is the maximal connected compact simple subgroup of the Lorentz group
$\Lor_{1,3}\equiv\SO_0(1,3)$, and the corresponding time-like energy--momentum
hyperboloid is $Y^3=\SO_3(1,3)/\SO(3)$. In this paper we show that in contrast
to this, for massless particles one obtains the line fixgroup $\Bor_{1,3}$ as
being the maximal connected noncompact solvable subgroup of the Lorentz group,
\[\begin{tabular}{cccccr}
$m\ne 0$&maximal&connected&compact&simple&{\bf point} fixgroup $\SO_3$\\
&$\updownarrow$&$\updownarrow$&$\times$&$\times$&\\
$m=0$&maximal&connected&noncompact&solvable&{\bf line} fixgroup $\Bor_{1,3}$
\end{tabular}\]
$\Bor_{1,3}\subset\Lor_{1,3}$ is the Borel subgroup of the Lorentz group, and
at least one of the manifolds $\Lor_{1,3}/\Bor_{1,3}$ is projective ($=S^2$).

Scanning the recent literature concerning Lie groups for massless particles,
few articles are found. Massless particles are considered within the
Poincar\'e group~\cite{Choi:2018mdd,Nistico:2021luz,Dragon:2024,%
Sazdovic:2024gaj,Sazdovic:2024wkz} but without any reference to the possible
solvability (solubility) of the corresponding stabilizer subgroup. Therefore,
our work in this field, in previous publications~\cite{Saar:2016jbx,%
Groote:2022} and in this publication, is pioneering and is based mainly on
handbooks. As the basis for a deeper understanding, we suggest
Refs.~\cite{Joos:1962qq,Niederer:1974ps,Niederer:1974pt,Borel:1991}, from
which many of the ideas that led to the research described in this paper were
derived. At this point, it is worth emphasizing that, as physicists, we do not
apply any rigorous mathematical formalism, consisting of definitions, lemmata,
prepositions, theorems, and corollaries, together with the corresponding
proofs. Instead, we mention the components of our considerations in these
mathematical terms, without providing proofs in most instances; these,
instead, can be found in the references that we cite at various points, which
also mark the ends of the corresponding theorems. An exception is
Lemma~\ref{lemma}, which is new and the main outcome of our analysis. This
lemma on the $\SL_2$ reconstruction of the proper Lorentz group can be
considered as a corollary of what has previously been discussed.

This publication is organized as follows. In Sec.~\ref{sec2}, we present
basic facts about the proper Lorentz group, and we present three theorems that
are the foundations of our work: the Chevalley theorem, the Lie--Kolchin
theorem, and a theorem related to Borel. Having explained Wigner's concept of
a little group, in Sec.~\ref{sec3} we first deal with Wigner's result for
this, given by the Euclidean group $E(2)$. However, the character equation has
an additional solution that leads to the Borel subgroup, dealt with in
Sec.~\ref{sec4}. In fact, there are a couple of these subgroups, with the
union giving the proper Lorentz group, while the cut is the maximal torus. We
show explicitly that each of these groups is generated by two elements of the
minimal solvable algebra $\sol_2$, spanning the algebra of the Borel subgroup
as a Kronecker sum. Turning to topology, in Sec.~\ref{sec5}, we show that
the quotient of the Lorentz group and the Borel subgroup is a projective
variety. Sec.~\ref{sec6} deals with representations, presenting also our
Lemma~\ref{lemma}, focused on reconstructing the proper Lorentz group by two
copies of the simplest noncompact group $\SL_2(\R)$. Via the Weinberg ansatz,
in Sec.~\ref{sec7} we describe a connection back to physics. In
Sec.~\ref{sec8} we present our conclusions.

\section{Basics\label{sec2}}
Here we give a brief exposition of the basic theorems and constructions of the
theory of semisimple groups with applications to the proper Lorentz group
\cite{Poincare:1905,Wigner:1939cj,Wigner:1962,Ohnuki:1976,Weinberg:1995,%
McRae:2025}. The Minkowski representation $\SO_0(1,3)$ preserves the
indefinite symmetric metric $\eta=\diag(1,-1,-1,-1)$ in the real spacetime
$\R^4\ni x,y$,
\[x\cdot y=x^\mu\eta_{\mu\nu} y^\nu=x^T\eta y.\]
As the defining representation of the causality-compatible Lorentz group,
\[\Lor_{1,3}=\{\Lambda^T\eta\Lambda=\eta,\ \det\Lambda=1,\ \Lambda^0{}_0\ge 1,\
  \Lambda\equiv(\Lambda^\mu{}_\nu)\in\GL_4(\R)\}\]
is parametrisable by six real parameters $\omega_{\mu\nu}=-\omega_{\nu\mu}$,
\[\Lambda(\omega)=\exp(-\frac12\omega_{\mu\nu}e^{\mu\nu})
  =\exp\left(\frac12\omega_p\epsilon^{0pjk}e_{jk}
  +\sum\omega_{0p}e_{0p}\right),\]
where the first part is compact and the second part noncompact, and the
domain of these six parameters is given by
\[D=\{\omega_{0p},\omega_p\in\R,\ -\pi<\omega_p\le\pi,\ p=1,2,3\}\]
which is homeomorphic to $\R^3\times\P_3$, where $\P_3$ is the
three-dimensional projective space. The generators have the form
\begin{equation}
e_{jk}=\begin{pmatrix}0&\vec 0^T\\
  \vec 0&\vec e_j\vec e_k^T-\vec e_k\vec e_j^T\end{pmatrix},\qquad
e_{0j}=\begin{pmatrix}0&\vec e_j^T\\\vec e_j&0_3\end{pmatrix},
\end{equation}
where in general $[e_{\mu\nu},e_{\rho\sigma}]=\eta_{\mu\rho}e_{\nu\sigma}
  +\eta_{\nu\sigma}e_{\mu\rho}-\eta_{\mu\sigma}e_{\nu\rho}
  -\eta_{\nu\rho}e_{\mu\sigma}$.

Here $\vec e_j$, $j=1,2,3$ is the Euclidean basis in $\R^3$. Therefore,
$\Lor_{1,3}$ is a locally compact and doubly connected, path-connected simple
and reductive group with universal covering $\SL(2,\C)$, i.e.\
\[\Lor_{1,3}\cong\SL(2,\C)/\Z_2\cong\SO(3,\C).\]
The last isomorphism means that the representations of $\Lor_{1,3}$ may be
seen as representations of the complex rotation group $\SO(3,\C)$. For
completeness, note that the Lie algebra $\lOR_{1,3}\equiv\log\Lor_{1,3}$ is
the noncompact real form of
\[\sO(4,\C)=\sL(2,C)\oplus\sL(2,\C).\]
To motivate what follows, it is instructive to look at Chevalley's theorem
in the context of the defining representation of the Lorentz group, according
to which $\Lor_{1,3}$ acts on the flat spacetime or its dual energy--momentum
space by a linear transformation,
\[\Lor_{1,3}\ni\Lambda:\E_{1,3}\ni x^\mu\mapsto\Lambda^\mu{}_\nu
  x^\nu\in\E_{1,3}.\]

\subsection{Chevalley theorem}
Let $G$ be a linear algebraic group and $H\subset G$ a closed algebraic
subgroup. Then there is a rational representation $\phi:G\to\GL(V)$ and a
one-dimensional subspace $L\subset V$ such that
\begin{equation}
H=\{g\in G:\phi(g)L=L\}.
\end{equation}
Otherwise, if $\ell\in L$ spans the line $L$, i.e.\ $L=\C\ell$, then the
equation
\begin{equation}\label{chareq}
\phi(h)\ell=\chi(h)\ell\in L,\ h\in H
\end{equation}
defines a character $\chi\in{\cal X}(H)$ of $H$. This character is called the
weight to the semi-invariant $\ell$, and ${\cal X}(H)=\Mor(H,\GL_1)$. Note
that if $[H,H]=H$, then ${\cal X}(H)=\{0\}$, i.e.\ $\SL(2,\C)$ has no
nontrivial characters and therefore the Lorentz group has no nontrivial
characters~\cite{Humphreys:1995}.

\subsection{Lie--Kolchin theorem}
Let $G$ be a connected solvable linear algebraic group, and let $(\phi,V)$ be
a regular representation of $G$. Then there exist characters
$\chi_i\in{\cal X}(G)$, $i=1,2,\ldots,n$ and a flag
\[V=V_1\supset V_2\supset\cdots\supset V_n\supset V_{n+1}=\{0\}\]
such that
\[(\phi(g)-\chi_i(g))V_i\subset V_{i+1}\]
for all $g\in G$. Taking $i=n$, one obtains the characteristic equation
\begin{equation}
\phi(g)V_n=\chi_n(g)V_n,
\end{equation}
i.e.\ every solvable group has a common one-dimensional subspace
$L\subset V$~\cite{Lie:1876,Kolchin:1948}.

\subsection{Wigner's little group}
What follows was proposed by Eugene Paul Wigner in 1939: for massive particles
($m\ne 0$) the point fixgroup (or little group) of the momentum
$\po=(1,0,0,0)$, ${\rm lg}(\po)=\SO(3)\subset\Lor_{1,3}$ is maximal connected,
compact and simple~\cite{Wigner:1939cj,Wigner:1962,Weinberg:1995,%
Sazdovic:2024gaj,Sazdovic:2024wkz}. Proceeding purely mathematically, one
finds that for massless particles ($m=0$), the little group is maximal
connected, noncompact and solvable, resulting in a Borel subgroup
$\Bor_{1,3}$. By definition, a Borel subgroup of an algebraic group $G$ is a
maximal connected solvable subgroup.

\subsection{Theorem}
Let $G$ be a connected linear algebraic group. Then
\begin{enumerate}
\item $G$ contains a Borel subgroup $B$.
\item All other Borel subgroups of $G$ are conjugate to $B$.
\item The homogeneous manifold $G/B$ is a projective variety.
\item $G=\bigcup_{g\in G}gBg^{-1}$, where $B$ is a fixed Borel subgroup of $G$.
\end{enumerate}
(see Theorem 11.4.7 in Ref.~\cite{GoodmanWallach}, p.~524).

So every element $g\in G$ is contained in a Borel subgroup. As mentioned
before, a solvable group has a semidirect structure. For the connected
solvable group $G$, the set $G_u$ of unipotent elements is a closed connected
nilpotent subgroup of $G$. There exists a maximal torus $T_G\subset G$ and for
this an exact sequence
\[e_G\to G_u\to G_u\rtimes T_G=G\to T_G\to e_G.\]
Since the maximal torus $T_G$ and the maximal connected unipotent subgroup
$G_u$ of $G$ are those for the Borel subgroup, one has the exact sequence
\begin{equation}
e_B\to B_u\to B_u\rtimes T_G=B\to T_G\to e_B.
\end{equation}
and $B=N_G(B_u)$. Here $G_u=\{g\in G:g=g_u\}$, where $g_u$ is the unipotent
component in the Jordan decomposition $g=g_sg_u$.

\section{The Role of Mass}\label{sec3}
Before discussing the Borel subgroup in further mathematical detail, let us
note some physical considerations regarding the kinematics. Taking the momentum
four-vector to be $p$, with $p^2=m^2>0$ the squared mass, for massive particles,
one can move to the rest frame, where $p=(m,0,0,0)^T$. The stabiliser subgroup
or fixgroup is given by $\SO(3)$, the three-dimensional rotations. However, if
the particle is massless, such a move to the rest frame is no longer possible.
According to Wigner's classification, the fixgroup is $E(2)$. This is the
little group that Wigner indicates for massless particles like photons and
(massless) neutrinos. For instance, for a momentum vector $p=(p_0,0,0,p_0)^T$
pointing in $z$ direction, $E(2)$ consists of rotations about the $z$ axis,
translations orthogonal to it and reflections. However, what is not taken
into account by this is the interchange of time and space components, which is
obviously an additional symmetry transformation. Together with this additional
transformation, the fixed point group is given by the Borel subgroup
$\Bor_{1,3}$.

Returning to mathematics, let $B\subset G$ be a Borel subgroup of $G$ and $V$
a finite-dimensional rational $G$-module. Then the fixed points of $B$ in $V$
coincide with the fixed points of $G$. As the Lorentz group has no fixed
points, for $B$, the character equation~(\ref{chareq}) is the only one that
can be solved. In contrast, for $\SO(3)$ one has the commutant
$[\SO(3),\SO(3)]=\SO(3)$. Therefore, the character group ${\cal X}(\SO(3))$
is trivial. The rotation group has no nontrivial characters, and the character
equation~(\ref{chareq}) is impossible to solve. Moreover, if $V$ is an
irreducible rotational $G$-module ($G$ be semisimple), then there is a
unique Borel-stable one-dimensional subspace spanned by a maximal vector of
some weight/character $\chi$ with multiplicity one.

The little group of Wigner is $E(2)$ which is nonmaximal, connected, solvable
and noncompact~\cite{Wigner:1939cj,Wigner:1962,Tung:1999,Croft:2025}. The
nonmaximality is explained (or determined) by the requirement to be a point
fixgroup of $(1,0,0,1)^T$,
\[E(2)\cong\R^2\rtimes\SO(2).\]
In this semidirect product, the compact group $SO(2)$ acts on the abelian
locally compact group $\exp\R^2$ by the multiplication rule
\[(x_1,R_1)(x_2,R_2)=(x_1+R_1x_2,R_1R_2).\]
Every irreducible representation of $E(2)$ is equivalent either to a
character of $\SO(2)$ lifted to $E(2)$, or to an induced representation
${\rm ind}_{\R^2}^{E(2)}\chi$, where $\chi$ is the nontrivial character of
$\R^2$. To avoid a continuum of helicity states, one has to require that for
physical states the noncompact part of $E(2)$ is trivial in all
representations, so the little group reduces to $\SO(2)$. There are
topological considerations that restrict the allowed values of the helicity
to integers or half-integers. Thus the helicity
\[\lambda=\frac{\vec J\vec k}{|\vec k\,|}=0,\pm\frac12,\pm 1,\ldots\]
is Lorentz invariant for massless particles with the total angular momentum
$\vec J$.

As for any Abelian group, the reducible representations of $\SO(2)$ are
one-dimensional. Therefore, according to Wigner's classification, the free
massless particles have only a single degree of freedom and are characterised
by the value $\lambda$ of their helicity.

In nature, there are two classes of particles. The first class consists of
particles that can exist in two helicity states $\pm\lambda$. Such a particle
is defined as a representation of the parity-extended Poincar\'e
group~\cite{Choi:2018mdd}. Since the electromagnetic interaction conserves
parity, the photon is defined as the $\SO(2)$-doublet
\[\left(\frac12,\frac12\right)\buildrel{\SO(2)}\over\longrightarrow
  (+1)\oplus(-1)\oplus 2\times(0)\buildrel{\rm parity}\over\longrightarrow
  (\pm 1)\oplus 2\times(0).\]
The second class contains particles for which the parity is not defined, as
the interactions they are involved in violate parity. Such particles are the
neutrinos that exist only with helicity $-\frac12$ and antineutrinos with
helicity $+\frac12$.

\section{The Borel group}\label{sec4}
The key observation in the preceding sections was the character
equation~(\ref{chareq}). Solving the character equation
\begin{equation}
B\po=\chi(B)\po
\end{equation}
for the light-like standard vector $\po=(1,0,0,1)$ with $B\in\Lor_{1,3}$, one
obtains~\cite{Saar:2016jbx,Groote:2022}
\begin{equation}
B^{(+)}(\vec\beta;\theta,\omega)=\begin{pmatrix}A&B^T\\\vec C&D\end{pmatrix}
\end{equation}
with
\begin{eqnarray}
A&=&\cosh\theta+\frac12|\vec\beta\,|^2e^{-\theta},\nonumber\\
\vec B&=&\left(\sinh\theta-\frac12|\vec\beta\,|^2e^{-\theta}\right)\vec e_3
  +\rot_3\omega\vec\beta,\nonumber\\
\vec C&=&\left(\sinh\theta+\frac12|\vec\beta\,|^2e^{-\theta}\right)\vec e_3
  +e^{-\theta}\vec\beta,\nonumber\\
D&=&\left(\cosh\theta-1-\frac12|\vec\beta\,|^2e^{-\theta}\right)\vec e_3
  \vec e_3^T-e^{-\theta}\vec\beta\vec e_3^T+(\1_3+\vec e_3\vec\beta^T)
  \rot_3\omega.
\end{eqnarray}
Here $\vec\beta^T=(\beta_1,\beta_2,0)=\sum_{a=1}^2\beta_a\vec e_a^T$,
$\vec e_3^T=(0,0,1)$,
\[\rot_3\omega=\begin{pmatrix}\cos\omega&-\sin\omega&0\\
  \sin\omega&\cos\omega&0\\0&0&1\end{pmatrix}.\]
As a consequence, the character is given by
\[\chi(B^{(+)})=\frac12\sum_{\mu,\nu=0,3}B^\mu{}_\nu=e^\theta.\]
The composition rule is
\begin{equation}
B^{(+)}(\vec\beta_1;\theta_1,\omega_1)B^{(+)}(\vec\beta_2;\theta_2,\omega_2)
  =B^{(+)}(\vec\beta_1+e^{\theta_1}\mbox{rot}_3\omega_1\vec\beta_2;
  \theta_1+\theta_2,\omega_1+\omega_2)
\end{equation}
and therefore
\begin{equation}
B^{(+)}(\vec\beta;\theta,\omega)
  =B_u^{(+)}(\vec\beta)\rtimes\Tra(\theta,\omega).
\end{equation}
All such transformations $B^{(+)}(\vec\beta;\theta,\omega)$ with noncompact
parameter space for helicity and gauge, given by $\{\vec\beta\in\R^2,\
\theta>0,\ 0<\omega\le\pi\}$, form the Borel subgroup
\begin{equation}
\Bor_{1,3}^{(+)}=(\Bor_{1,3}^{(+)})_u\rtimes\Tor_{1,3}\subset\Lor_{1,3}.
\end{equation}
Here $\Tor_{1,3}=\SO_0(1,1)\times\SO(2)$ is the maximal torus in $\Lor_{1,3}$,
and $(\Bor_{1,3}^{(+)})_u$ is the unipotent radical of $\Bor_{1,3}^{(+)}$.

The linearisation of $\Bor_{1,3}^{(+)}$ in the neighborhood of the identity,
\begin{equation}
B^{(+)}(\vec\beta;\theta,\omega)=\1_4+\beta_1b_1+\beta_2b_2+\theta b_0
  +\omega b_3=\1_4-\frac12\omega_{\mu\nu}e^{\mu\nu}
\end{equation}
results in the Lie algebra $\bor_{1,3}^{(+)}=\log\Bor_{1,3}^{(+)}$, with
\begin{equation}
b_0=e_{03},\quad b_1=e_{01}+e_{31},\quad b_2=e_{02}+e_{32},\quad b_3=e_{21}.
\end{equation}
As a vector space $\kest_\R\{b_\mu\}_0^3$ endowed with commutation relations
\begin{equation}
[b_0,b_a]=b_a,\qquad[b_3,b_a]=-\epsilon_{3ab}b_b,\qquad a=1,2
\end{equation}
the Borel algebra reads
\begin{equation}
\bor_{1,3}^{(+)}=(\bor_{1,3}^{(+)})_u\rtimes\tor_{1,3}
  =\sO(1,1)\oplus\sO(2)\oplus\R^2,
\end{equation}
where $(\bor_{1,3}^{(+)})_u=\rad_u^{(+)}$ is the Lie algebra corresponding to
the unipotent radical.

Since the semisimple rank is $\rank_{\rm ss}\SO(1,3)=2$, there exists a unique
Borel subgroup $\Bor_{1,3}^{(-)}\subset\Lor_{1,3}$, called opposite
$\Bor_{1,3}$, such that~\cite{Humphreys:1995,Shaw:1964zz,Bump:2004,Saller:2010}
\[\Bor_{1,3}^{(-)}\cap\Bor_{1,3}^{(+)}=\Tor_{1,3}=\SO_0(1,1)\times\SO(2)\]
and $\Lor_{1,3}=\Bor_{1,3}^{(-)}\cup\Bor_{1,3}^{(+)}$.

To see the algebraic structure of $\bor_{1,3}^{(+)}$, it is convenient to
transform the basis to
\begin{eqnarray}
t_0&=&\frac12(b_0+ib_3),\quad t_+\ =\ \frac12(ib_1-b_2),\qquad
  [t_0,t_+]\ =\ t_+\\
u_0&=&\frac12(b_0-ib_3),\quad u_+\ =\ \frac12(-ib_1-b_2),\qquad
  [u_0,u_+]\ =\ u_+
\end{eqnarray}
with $[t_{0,+},u_{0,+}]=0$. This leads to the Kronecker sum decomposition
\begin{equation}
\bor_{1,3}^{(+)}=\sol_2(e)\boxplus\sol_2(f)
  :=\sol_2(e)\otimes\1_2+\1_2\otimes\sol_2(f).
\end{equation}
The Kronecker sum decomposition is easily seen after applying the splitting
map,
\begin{eqnarray}
t_0=\frac12h\otimes\1_2,&&t_+=(ie)\otimes\1_2\qquad\mbox{for } \sol_2(e),
  \nonumber\\
u_0=\1_2\otimes(-\frac12h),&&u_+=\1_2\otimes(if)\qquad\mbox{for } \sol_2(f).
\end{eqnarray}
Here
\[h=\begin{pmatrix}1&0\\0&-1\end{pmatrix},\qquad
e=\begin{pmatrix}0&1\\0&0\end{pmatrix},\qquad
f=\begin{pmatrix}0&0\\1&0\end{pmatrix}\]
is the natural Chevalley basis of $\sL_2(\C)$.

Again, the linearisation of $\Bor_{1,3}^{(-)}$ generates the basis of the
underlying vector space as
\begin{equation}
k_0=-e_{03},\quad k_3=e_{21},\quad k_a=-e_{0a}+e_{3a},\quad a=1,2
\end{equation}
with nonzero commutation relations
\begin{equation}
[k_0,k_a]=k_a,\qquad[k_3,k_a]=-\epsilon_{3ab}k_b.
\end{equation}
So
\[\lOR_{1,3}=\bor_{1,3}^{(-)}+\bor_{1,3}^{(+)}=(\bor_{1,3}^{(-)})_u
  \oplus(\bor_{1,3}^{(+)})_u+\tor_{1,3},\]
where $\tor_{1,3}=\log\Tor_{1,3}=\log\SO_0(1,1)\oplus\log\SO(2)$.

The Kronecker sum decomposition of $\bor_{1,3}^{(+)}$ into two fundamental
solvable groups $\sol_2$ is easily extended to an $\sL_2(\R)$ decomposition
for the whole $\lOR_{1,3}$,
\[\lOR_{1,3}\cong\sL_2(\R)_e\boxplus\sL_2(R)_f.\]
Using the splitting map, for $\sL_2(\R)_e$ one obtains
\begin{eqnarray}
t_0&=&\frac12(b_0+ib_3)\ \to\ \left(\frac12h\right)\otimes\1_2\nonumber\\
t_+&=&\frac12(ib_1-b_2)\ \to\ (ie)\otimes\1_2\\
t_-&=&\frac12(-ik_1-k_2)\ \to\ (if)\otimes\1_2\nonumber
\end{eqnarray}
with commutation relations
\[[t_+,t_-]=-2t_0,\qquad[t_0,t_\eps]=\eps t_\eps,\quad\eps=\pm\]
For $\sL_2(\R)_f$ one obtains
\begin{eqnarray}
u_0&=&\frac12(b_0-ib_3)\ \to\ \1_2\otimes\left(-\frac12h\right)\nonumber\\
u_+&=&\frac12(-ib_1-b_2)\ \to\ \1_2\otimes(if)\\
u_-&=&\frac12(ik_1-k_2)\ \to\ \1_2\otimes(ie)\nonumber
\end{eqnarray}
with commutation relations
\[[u_+,u_-]=-2t_0,\qquad[u_0,u_\eps]=\eps t_\eps,\quad\eps=\pm\]
and $[\sL_2(\R)_e,\sL_2(\R)_f]=0$.

Returning from the defining matrix representation of the Borel algebra to the
group matrix representation, by exponentiation one obtains
\[\tor_{1,3}\ni\vartheta b_0+\omega b_3\buildrel{\exp}\over\longrightarrow
  \exp(\vartheta b_0+\omega b_3)\in\Tor_{1,3}\]
with
\begin{eqnarray}
\exp(\vartheta b_0+\omega b_3)&=&\left(\1_4+\sinh\vartheta b_0
  +(\cosh\vartheta-1)b_0^2\right)\left(\1_4+\sin\omega b_3
  +(1-\cos\omega)b_3^2\right)\ =\nonumber\\
  &=&\begin{pmatrix}\sqrt{1+\vec x^T\vec x}&\vec x^T\\
  \vec x&\rot_3\omega\sqrt{\1_3+\vec x\vec x^T}\end{pmatrix},\qquad
  \vec x=\sinh\vartheta\vec e_3.\nonumber
\end{eqnarray}
The Cartan--Killing form in the defining representation is indefinite,
\[(\vartheta b_0+\omega b_3,\vartheta b_0+\omega b_3)
  =2(\vartheta^2-\omega^2)\]
and
\[\left(\exp(\vartheta b_0+\omega b_3),\exp(\vartheta b_0+\omega b_3)
  \right)=2\left(\cosh\vartheta+\cos\omega\right)\ge 0.\]
For the unipotent radical one has
\[\rad_u^{(+)}\ni b =\beta_1b_1+\beta_2b_2
  \buildrel{\rm exp}\over\longrightarrow
  \exp b=-\1_4+\exp(\beta_1b_1)+\exp(\beta_2b_2).\]
Since the algebra $\bor_{1,3}^{(+)}$ is solvable, and its derived algebra is
given by
\[\Der\bor_{1,3}^{(+)}=\rad_u^{(+)},\]
the Cartan--Killing form is identically zero,
$(\rad_u^{(+)},\rad_u^{(+)})\equiv 0$. For $\Rad_u^{(+)}$ one obtains
$(\exp b,\exp b)=4$, so obviously $b\in\rad_u^{(+)}$ nilponent and
$\exp b\in\Rad_u^{(+)}$ unipotent.

\section{The quotients}\label{sec5}
Given a closed subgroup $H$ of an algebraic group $G$, there is a smooth
projection $\pi:G\to G/H$, where the fibres are precisely the cosets $gH$.
The projection $\pi$ has a smooth local injection, given by the compatible
section
\begin{center}\begin{tabular}{ccc}
  $\gamma:G/H$&$\rightarrow$&$G$\\
  &$\llap{id}\searrow$&$\downarrow\rlap{$\pi$}$\\
  &&$G/H$
\end{tabular}\end{center}
such that $\pi\circ\gamma=\id_{G/H}$.

As an example, we consider the Cartan decomposition
$\mathfrak g=\log G=\log K\oplus\mathfrak p$, where $\log K$ is the maximal
compact subalgebra of $\mathfrak g$ and the subspace
$\mathfrak p=\mathfrak g\mod\log K$ consists of the noncompact generators of
$\mathfrak g$. Exponentiating the Lie algebra decomposition into the Lie group,
\[\begin{tabular}{lccc}
$\mathfrak g=$&$\log K$&$\oplus$&$\mathfrak p$\\
&compact&&noncompact\\
$\llap{$\exp$}\downarrow$&$\downarrow$&&$\downarrow$\\
$G=$&$K$&$\otimes$&$\exp\mathfrak p$\\
&compact&&coset\\
&subgroup&&representatives
\end{tabular}\]
one obtains the parametrisation of the algebraic manifold
$G/K=\exp\mathfrak p$.

The map of $K\times\mathfrak p$ onto $G$,
\[K\times\mathfrak p\ni(k,X)\to k\exp X\in G\]
is a diffeomorphism into $G$, i.e.\ $G=K\times\exp\mathfrak p$.

Different choices of the section $\gamma$ give different formulae for the
coset representatives. For the Borel subgroup $B\subset G$, the factor set
$G/B$ is the largest homogeneous space for $G$ having the structure of a
projective variety. Since $G/B$ is complete, the Borel subgroup $B$ has a
fixed point in $G/B$.

\subsection{The $\SO(3)$ parametrisation}
The Borel decomposition of the Lorentz group $\Lor_{1,3}=\SO_0(1,3)$ is
generated by the decomposition of the algebra $\lOR_{1,3}=\log\SO_0(1,3)$ in
a natural way by reordering the usual parametrisation
\[\begin{tabular}{rccc}
$\lOR_{1,3}\ni-\frac12\omega_{\mu\nu}e^{\mu\nu}=$&
$\underbrace{\omega_1e_{31}+\omega_2e_{32}}$&$+$&
$\underbrace{\beta_1b_1+\beta_2b_2+\vartheta e_{03}+\omega e_{21}}$\\
&coset representatives&&$\bor_{1,3}$\\
&$\llap{$\exp$}\downarrow$&&$\downarrow$\\
$\Lor_{1,3}\ni$&$\exp\sum_{a=1}^2\omega_ae_{3a}$&$\times$&
$\Bor_{1,3}^{(+)}(\beta_1,\beta_2;\vartheta,\omega)$
\end{tabular}\]
More precisely,
\begin{eqnarray}
\lefteqn{\P_{(2)}\ \equiv\ \sum_{a=1}^2\omega_ae_{3a}\ =\ \begin{pmatrix}
  0&\vec 0^T\\\vec 0&\vec e_3\vec\omega_{(3)}^T-\vec\omega_{(3)}\vec e_3^T
  \end{pmatrix}}\nonumber\\
  &\to&\exp\P_{(2)}=\begin{pmatrix}1&\vec 0^T&0\\
  \vec 0&\sqrt{\1_2-\vec x_{(2)}\vec x_{(2)}^T}&-\vec x_{(2)}\\
  0&\vec x_{(2)}^T&\sqrt{1-\vec x_{(2)}^T\vec x_{(2)}}\end{pmatrix}.\nonumber
\end{eqnarray}
Here
\[\vec x_{(2)}=\frac{\sin\omega_{(2)}}{\omega_{(2)}}\vec\omega_{(2)},\quad
  \vec\omega_{(2)}=\begin{pmatrix}\omega_1\\\omega_2\end{pmatrix},\quad
  \omega_{(2)}^2=\omega_1^2+\omega_2^2,\quad
  \vec\omega_{(3)}=\begin{pmatrix}\vec\omega_{(2)}\\0\end{pmatrix}.\]

The Cartan--Killing inner product for representatives
\[(\P_{(2)},\P_{(2)})=\tr\P_{(2)}^2=-2\omega_{(2)}^2<0\]
is negative, i.e.\ the representatives are compact operators. For the
representatives of the group coset one has
\[(\exp\P_{(2)},\exp\P_{(2)})=4\cos^2\omega_{(2)}>0.\]
Therefore, the geometric manifold for the respresentatives of the group coset
is compact. Moreover, the parametrisation of the $\Bor_{1,3}$-classes can be
given by the three-point $\vec x=(x_1,x_2,x_3)^T$ in $\R^3$ as
\[\vec x_{(2)}=\frac{\sin\omega_{(2)}}{\omega_{(2)}}\vec\omega_{(2)}
  =\begin{pmatrix}x_1\\x_2\end{pmatrix}\]
and $x_3=\cos\omega_{(2)}$. Then
\[\exp\P_{(2)}=\begin{pmatrix}1&\vec 0^T&0\\
  \vec 0^T&\displaystyle\1_2-\frac1{1+x_3}\vec x_{(2)}\vec x_{(2)}^T
  &-\vec x_{(2)}\\0&\vec x_{(2)}^T&x_3\end{pmatrix}\]
with $\det\exp\P_{(2)}=x_1^2+x_2^2+x_3^2=1$. Therefore, the $\SO$-type coset
representatives generate a compact factor set, the two-sphere in
$\R^3$~\cite{VilelaMendes:2025mua},
\[\Lor_{1,3}/\Bor_{1,3}^{(+)}=\SO(3)/\SO(2)=S^2.\]
Using the Iwasawa decomposition
\begin{eqnarray}
\Lor_{1,3}&=&\SO(3)\times\SO_0(1,1)\times\exp\R^2\nonumber\\
  &=&(\SO(3)/\SO(2))\times
  \underbrace{\SO(2)\times SO_0(1,1)\times\exp\R^2}_{\Bor_{1,3}}\nonumber
\end{eqnarray}
and $\Bor_{1,3}^{(+)}=(\SO(2)\times\SO_0(1,1))\ltimes\exp\R^2$, one obtains
the same result.

The projective coordinates for this parametrisation are
\[-\infty<z_a=\frac{x_a}{\sqrt{1-x_1^2-x_2^2}}=\frac{x_a}{x_3}<\infty,\quad
a=1,2.\]
As mentioned, different choices for the section $\gamma$ provide
different parametrisations.

\subsection{The $\SO(1,2)$ parametrisation}
Let
\begin{center}\begin{tabular}{ccc}
  $\llap{$\lOR_{1,3}\ni-\frac12\omega_{\mu\nu}e^{\mu\nu}=\ $}%
  \vartheta_1e_{01}+\vartheta_2e_{02}$&$+$
  &$\beta_1b_1+\beta_2b_2+\vartheta e_{03}+\omega e_{21}$\\
  coset representatives&&$\bor_{1,3}^{(+)}$\\
  $\llap{$\exp$}\downarrow$&&$\downarrow$\\
  $\llap{$\Lor_{1,3}\ni$}\exp\vartheta_{(2)}$&$\times$&$\Bor_{1,3}^{(+)}$
  \end{tabular}\end{center}
Here $\vartheta_{(2)}=\sum_{a=1}^2\vartheta_ae_{0a}$ is noncompact, as the
Cartan--Killing form is positive,
\[(\vartheta_{(2)},\vartheta_{(2)})=2\vartheta_{(2)}^2
  =2(\vartheta_1^2+\vartheta_2^2)>0.\]
The group representatives $\exp\vartheta_{(2)}$ are noncompact,
\[\SO_0(1,2)\ni\exp\vartheta_{(2)}
  =\begin{pmatrix}\sqrt{1+\vec y_{(3)}^T\vec y_{(3)}}&\vec y_{(3)}^T\\
  \vec y_{(3)}&\sqrt{\1_3+\vec y_{(3)}\vec y_{(3)}^T}\end{pmatrix},\quad
  \vec y_{(3)}=\frac{\sinh\vartheta_{(2)}}{\vartheta_{(2)}}
  \begin{pmatrix}\vartheta_1\\\vartheta_2\\0\end{pmatrix},\]
as the Cartan--Killing form is positive,
\[(\exp\vartheta_{(2)},\exp\vartheta_{(2)})=4\cosh^2\vartheta_{(2)}>0.\]
The parametrisation of the coset representations for the group can be chosen
as the point $y=(y_0,y_1,y_2)$ in $\R^3$, or as the projective coordinates
$z_a=y_a/y_0$, $a=1,2$ in the interior of the unit circle of the $y_0=1$
plane. In the first case, we have
\[y_0=\cosh\vartheta_{(2)},\quad
y_a=\frac{\sinh\vartheta_{(2)}}{\vartheta_{(2)}}\vartheta_a,\quad a=1,2,\]
so that $y_0^2-y_1^2-y_2^2=1$, and the representatives are
\[Q(y_0,y_1,y_2)=\begin{pmatrix}y_0&\vec y_{(3)}^T\\
  \vec y_{(3)}&\displaystyle\1_3+\frac1{1+y_0}\vec y_{(3)}\vec y_{(3)}^T
  \end{pmatrix}\]
with $\det Q=(y_0^2-y_1^2-y_2^2)/(1+y_0)=1$ and $(Q,Q)=4y_0^2>0$. As a
consequence, the representatives are found on the noncompact on-shell
hyperboloid
\[Y^2=\Lor_{1,3}/\Bor_{1,3}^{(+)}=\SO_0(1,2)/\SO(2).\]
The projective coordinates $(q_1,q_2)$ are
\[-1<q_a=\frac{y_a}{\sqrt{1+y_1^2+y_2^2}}=\frac{y_a}{y_0}<1,\quad a=1,2.\]

Therefore, the two examples for the coset representatives considered up to
now are the time-like on-shell hyperboloid $Y^2$ and (as compact partner)
the sphere $S^2$ with the common compact subgroup given by
$\SO(2)\subset\Tor_{1,3}$.

\subsection{The Borel parametrisation}
Finally, to be systematic, the Borel structure of the Lorentz group provides
a constructive procedure to determine representatives of different cosets.
To begin with, we recall the Borel decomposition
\[\lOR_{1,3}=\bor_{1,3}^{(-)}\cup\bor_{1,3}^{(+)}
  =\rad_u^{(-)}\oplus(\rad_u^{(+)}\rtimes\tor_{1,3})
  =\rad_u^{(+)}\oplus(\rad_u^{(-)}\rtimes\tor_{1,3}).\]
  As underlying vector spaces one has
\[\vec\lOR_{1,3}=\kest_\R\{e_{\mu\nu}=-e_{\nu\mu}\}_0^3
  =\kest_\R\{b_0,b_3,b_a,k_a\}_{a=1}^2
  =\kest_\R\{t_0,t_\eps;u_0,u_\eps\}_{\eps=\pm 1}.\]
It is convenient to choose the basis of $\rad_u^{(-)}$ as
\[k_a=\begin{pmatrix}0&-\vec e_a^T\\-\vec e_a
  &\vec e_3\vec e_a^T-\vec e_a\vec e_3^T\end{pmatrix}.\]
Then
\[\rad_u^{(-)}\ni k=\kappa_1k_1+\kappa_2k_2=\begin{pmatrix}
  0&-\vec\kappa_{(3)}^T\\-\vec\kappa_{(3)}
  &\vec e_3\vec\kappa_{(3)}^T-\vec\kappa_{(3)}\vec e_3^T\end{pmatrix},\]
where
\[\vec\kappa_{(3)}=\begin{pmatrix}\kappa_1\\\kappa_2\\0\end{pmatrix},\quad
\kappa_{(3)}^2=\kappa_1^2+\kappa_2^2,\]
and the Cartan--Killing product $(k,k)=\tr k^2=0$. This is the
second criterion for the solvability of $\bor_{1,3}^{(-)}$: an algebra
$\mathfrak g$ is solvable if and only if its Cartan--Killing metric tensor is
identically zero on its derived algebra ${\cal D}^1\mathfrak g$.

The expression for the representatives of the group coset is obtained by
exponentiation,
\[\begin{tabular}{rccl}
$\lOR_{1,3}\supset$&$\rad_u^{(-)}$&$\oplus$&$\bor_{1,3}^{(+)}$\\
&$\llap{$\exp$}\downarrow$&&$\downarrow$\\
$\Lor_{1,3}\supset$&$\Rad_u^{(-)}$&$\times$&$\Bor_{1,3}^{(+)}$
\end{tabular}\]
From this, the representations of the group coset can be read off as
\[\Rad_u^{(-)}\ni\exp k=\1_4+k+\frac12k^2=\begin{pmatrix}
1+t&-x&-y&t\\-x&1&0&-x\\-y&0&1&-y\\-t&x&y&1-t\end{pmatrix},\]
where $t=\frac12\kappa_{(3)}^2$, $x=\kappa_1$, $y=x_2$ and $2t-(x^2+y^2)=0$.
One observes that the real parameters $(t,x,y)$ describe the subspace of the
noncompact elliptic paraboloid. The projective coordinates of this
parametrisation are
\[(p_1,p_2)=\left(\frac xt,\frac yt\right)
  =\left(\frac x{x^2+y^2},\frac y{x^2+y^2}\right),\quad p_1^2+p_2^2>0.\]
The Cartan--Killing form for the representatives of the group coset is
positive,
\[(\exp k,\exp k)=4>0.\]
As a consequence, one has the following:
\begin{itemize}
\item[$m\ne 0$:] a rest frame exists for massive particles $\Rightarrow$\\
  stabiliser subgroup is the point fixgroup $\SO(3)$ $\Rightarrow$ leading to
  spin,
\item[$m=0$:] no rest frame exists for massless particles $\Rightarrow$\\
  stabiliser subgroup is the line fixgroup $\Bor_{1,3}$ $\Rightarrow$
  leading to helicity.
\end{itemize}
Indeed, accepting the undulatory theory of light, the plane wave as the most
elementary type of wave cannot be localised in space. Moreover, the
characteristic equation~(\ref{chareq}) suggests that the massless particle
can be enclosed on a line. Therefore, the question of how a massless particle
with energy $E$ differs from the same particle with energy $\lambda E$ is a
quantum mechanical problem in the form of Planck's formula $E=h\nu$, rather
than a problem of symmetry. In fact, the difference is a mathematical one,
\begin{itemize}
\item[$m\ne 0$:] semisimple compact,
\item[$m=0$:] solvable noncompact.
\end{itemize}

\section{The representations}\label{sec6}
In general, Lie algebras play their role in physics not as abstract algebras
but through their representations that act on suitable representation spaces.
For example, spin and helicity are determined by the stabiliser subgroups of
$\Lor_{1,3}$. For mathematical convenience it is reasonable to consider
representations in vector spaces over complex number fields. This is because
in physics the concept of reducibility is of fundamental importance, and the
mathematical structure of quantum mechanics works with complex Hilbert
spaces~\cite{Wigner:1939cj,Wigner:1962,Ohnuki:1976,Weinberg:1995,Tung:1999,%
Wightman:1973,Bauerle:1990,Barut:1986dd}.

A representation $D$ of a real algebra $\mathfrak g$ can be extended to a
unique linear complex representation $\hat D$ by the holomorphic extension
\[\hat D(A+iB)=D(A)+iD(B),\quad A,B\in\mathfrak g.\]
Although $\Lor_{1,3}\cong\SO(3,\C)$, the process of holomorphic extension for
$\lOR_{1,3}$ arises from the fact that $\lOR_{1,3}$ is the real form of
$\sO_4(\C)$,
\[\lOR_{1,3}\to\C\otimes_\R\lOR_{1,3}\cong\sO_4(\C)
  =\sL(2,\C)\oplus\sL(2,\C)=\C\otimes_\R\su(2)\oplus\C\otimes_\R\su(2).\]
This complexification of $\lOR_{1,3}$ provides the link between the
real-valued Lorentz algebra $\lOR_{1,3}$ and the real-valued algebra $\su(2)$,
and using this link, one can construct all the representations of $\lOR_{1,3}$.

Since a representation $\tau_m$ of $\sO(3)$ comes from a representation
$D^{(m)}$ of $\SO(3)\subset\Lor_{1,3}$, the diagram
\[\begin{tabular}{ccc}
$\tau_m:\ \sO_3\in x$&$\longrightarrow$&$\tau_m(x)$\\
$\llap{$\exp$}\downarrow$&&$\downarrow$\\
$T^{(m)}:\ \SO(3)\in\exp x$&$\longrightarrow$&$D^{(m)}(\exp x)=\exp\tau_m(x)$
\end{tabular}\]
has to be commutative. However for odd $m$ there is no such representation
$D^{(m)}$. In order to overcome this problem, one needs $\su(2)$ to generate
all the finite-dimensional representations of $\lOR_{1,3}$. The complex
representations $T^{(k,l)}$ ($k,l=0,\frac12,1,\ldots$) may be obtained by
holomorphic extension and Weyl's unitary trick,
\begin{eqnarray}
\lefteqn{\{e_{\mu\nu}\}_0^3\ \xrightarrow[\rm unitary\ trick]{\rm Weyl's}\
  \left\{\frac12\left(-\frac12\epsilon_{pnq}e_{nq}\pm ie_{0p}\right)
  \right\}_1^3}\\\strut&\xrightarrow[\rm map]{\rm splitting}&
  \left\{D_p=m_p\boxplus m_p,\ B_p=-i(m_p\boxplus(-m_p))\right\}_1^3\nonumber\\
  \strut&\xrightarrow{T^{(k,l)}}&\kern-12pt\left\{T^{(k,l)}(D_p)
  =D^{(k)}(m_p)\boxplus D^{(l)}(m_p),\ T^{(k,l)}(B_p)
  =-i\left(D^{(k)}(m_p)\boxplus(-D^{(l)}(m_p))\right)\right\}_1^3.\kern-1pt
  \nonumber
\end{eqnarray}
Here $\{m_p\}_1^3$ generate the algebra $\su(2)$ and $D^{(k)}$,
$k=0,\Frac12,1,\ldots$ are the common representations of $\su(2)$.

Any irreducible finite-dimensional representation of $\lOR_{1,3}$ is
isomorphic to $T^{(k,l)}$ for some $(k,l)$. As a special case, there are two
inequivalent fundamental representations from which all others can be obtained
by reducing the tensor products. The two-dimensional spinor representation
$(\frac12,0)$ is defined by the commutative diagram
\[\begin{tabular}{rccl}
$\Lor_{1,3}\ni\Lambda$:&$E_{1,3}\ni p$&$\Longrightarrow$&$\Lambda p$\\
$\llap{$T^{(1/2,0)}$}\downarrow$&$\llap{$\sigma$}\downarrow$&
  &$\llap{$\sigma$}\downarrow$\\
$\SL_2(\C)\ni T^{(1/2,0)}(\Lambda)\equiv A_\Lambda$:
  &$\H_2\ni\sigma(p)$&$\longrightarrow$&$A_\Lambda\sigma(p)A_\Lambda^\dagger
  =\sigma(\Lambda p)$\end{tabular}\]
The representation $(0,\frac12)$ is defined as
\[\begin{tabular}{rccl}
$\Lor_{1,3}\ni\Lambda$:&$E_{1,3}\ni p$&$\Longrightarrow$&$\Lambda p$\\
$\llap{$T^{(0,1/2)}$}\downarrow$&$\llap{$\tilde\sigma$}\downarrow$&
  &$\llap{$\tilde\sigma$}\downarrow$\\
$\SL_2(\C)\ni T^{(0,1/2)}(\Lambda)\equiv\tilde A_\Lambda$:
  &$\H_2\ni\tilde\sigma(p)$&$\longrightarrow$
  &$\tilde A_\Lambda\tilde\sigma(p)\tilde A_\Lambda^\dagger
  =\tilde\sigma(\Lambda p)$\end{tabular}\]
Here $\tilde\sigma(p)=p^\mu\tilde\sigma_\mu$, $\sigma(p)=p^\mu\sigma_\mu$,
with $\tilde\sigma_0=\sigma_0=\1_2$ and $\tilde\sigma_p=-\sigma_p$ ($p=1,2,3$)
the Pauli matrices. Since the finite-dimensional representations of $\su(2)$
are in one-to-one correspondence with those of $\sL_2(\R)$ and the Lie algebra
$\lOR_{1,3}$ is noncompact, it is reasonable to define the representations of
$\lOR_{1,3}$ in terms of the algebra $\sL_2(\R)$. Moreover, the most important
technique to study the representations of linear noncompact groups is to
reduce the problem to the subgroups isomorphic to the simplest noncompact
group $\SL_2(\R)$. For example, $\lOR_{1,3}$ contains three such
$\sL$-isomorphic subalgebras but only one $\su$-isomorphic subalgebra. The
following lemma gives the $\sL$-structure for $\lOR_{1,3}$.

\subsection{Lemma}\label{lemma}
If $e_{\mu\nu}=-e_{\nu\mu}$, $\mu,\nu=0,1,2,3$ are defined by
\begin{eqnarray}
e_{01}&=&e_2\boxplus(-e_2)\nonumber\\
e_{02}&=&-i(e_1\boxplus(-e_1))\nonumber\\
e_{03}&=&e_3\boxplus(-e_3)\nonumber\\
e_{31}&=&e_1\boxplus e_1\nonumber\\
e_{32}&=&-i(e_2\boxplus e_2)\nonumber\\
e_{21}&=&-i(e_3\boxplus e_3),\nonumber
\end{eqnarray}
where $\{e_k\}_1^3$ are the generators of the algebra $\sL_2(\R)$,
\[[e_3,e_1]=e_2,\quad[e_3,e_2]=e_1,\quad
  [e_1,e_2]=-e_3,\]
then $e_{\mu\nu}$ generate the Lorentz algebra $\lOR_{1,3}$,
\[[e_{\mu\nu},e_{\rho\sigma}]=\eta_{\mu\rho}e_{\nu\sigma}
  +\eta_{\nu\sigma}e_{\mu\rho}-\eta_{\mu\sigma}e_{\nu\rho}
  -\eta_{\nu\rho}e_{\mu\sigma}.\]

Applying this lemma, one can define the representations of $\lOR_{1,3}$ in
terms of the holomorphic extensions of the irreducible representations
$\pi^{(n)}$ of $\sL_2(\R)$~\cite{Shaw:1964zz,Bump:2004,Saller:2010,%
Campoamor-Stursberg:2024fpl},
\begin{eqnarray}
\pi^{(k,l)}(e_{01})&=&\pi^{(k)}(e_2)\boxplus\left(-\pi^{(l)}(e_2)
  \right)\nonumber\\
\pi^{(k,l)}(e_{31})&=&\pi^{(k)}(e_1)\boxplus\pi^{(l)}(e_1)\\
\pi^{(k,l)}(e_{21})&=&-i\left(\pi^{(k)}(e_3)\boxplus\pi^{(l)}(e_3)
  \right)\nonumber
\end{eqnarray}
($k,l=0,\frac12,1,\ldots$). Here $\pi^{(k)}$ is the standard representation of
$\sL_2(\R)$,
\begin{eqnarray}\label{pik123}
\pi^{(k)}(e_1)|k,m\rangle
  &=&\frac12\rho_m^{(k)}|k,m+1\rangle-\frac12\rho_{m-1}^{(k)}|k,m-1\rangle,
  \nonumber\\
\pi^{(k)}(e_2)|k,m\rangle
  &=&\frac12\rho_m^{(k)}|k,m+1\rangle+\frac12\rho_{m-1}^{(k)}|k,m-1\rangle,
  \nonumber\\
\pi^{(k)}(e_3)|k,m\rangle&=&m|k,m\rangle,
\end{eqnarray}
with $\rho_m^{(k)}=\sqrt{(k+m+1)(k-m)}$, $m=-k,-k+1,\ldots,k$.

In some contexts, it is more convenient to work with the $\sL$-basis
$e=e_1+e_2$, $f=-e_1+e_2$, and $h=2e_3$. By the lemma, the Borel algebras have
the form
\begin{eqnarray}\label{decompb}
b_0&=&\frac12(h\boxplus(-h))\nonumber\\
b_1&=&e\boxplus(-f)\nonumber\\
b_2&=&-i(e\boxplus f)\nonumber\\
b_3&=&-\frac i2(h\boxplus h)
\end{eqnarray}
for $\bor_{1,3}^{(+)}$ and
\begin{eqnarray}\label{decompk}
k_0&=&-\frac12(h\boxplus(-h))\nonumber\\
k_1&=&(-f)\boxplus e\nonumber\\
k_2&=&-i(f\boxplus e)\nonumber\\
k_3&=&-\frac i2(h\boxplus h)
\end{eqnarray}
for the opposite $\bor_{1,3}^{(-)}$.

The $(2k+1)$-dimensional representation $\pi^{(k)}$ of $\sL_2(\C)$ is
\begin{eqnarray}
\pi^{(k)}(h)|k,m\rangle&=&2m|k,m\rangle\nonumber\\
\pi^{(k)}(e)|k,m\rangle&=&\rho_m^{(k)}|k,m+1\rangle\nonumber\\
\pi^{(k)}(f)|k,m\rangle&=&\rho_{m-1}^{(k)}|k,m-1\rangle
\end{eqnarray}
with $\rho$ and $k$ as in Eq.~(\ref{pik123}).

\subsection{Theorem}
Let $2k\in\N$ and let $(\pi,V)$ be a simple representation of $\sL_2(\C)$ of
dimension $2k+1$. Then
\begin{enumerate}
\item $\pi$ is equivalent to $\pi^{(k)}$ for some $k$
\item the eigenvalues of $\pi^{(k)}(h)$ are
  $\{-2k,-2k-2,\ldots,2k\}=\Spec\pi^{(k)}(h)$
\item if $0\ne v\in V$ satisfies $\pi^{(k)}(e)v=0$, then $\pi^{(k)}(h)v=2kv$,
  i.e.\ $\pi^{(k)}(h)$ and $\pi^{(k)}(e)$ have the common eigenvector
  $|k,k\rangle$.
\item if $0\ne v\in V$ satisfies $\pi^{(k)}(f)v=0$, then $\pi^{(k)}(h)v=-2kv$,
  i.e.\ $\pi^{(k)}(h)$ and $\pi^{(k)}(f)$ have the common eigenvector
  $|k,-k\rangle$.
\end{enumerate}
(Theorem~19.2.5 in Ref.~\cite{TauYu}, p.~281).

As a matter of fact, the statements~3 and~4 generate/define the eigenvectors
of the representation $\pi^{(k,l)}$, called the helicity states for
$\bor_{1,3}$. Using the $\sL$-decomposition~(\ref{decompb}) of $\bor_{1,3}$,
one obtains
\begin{eqnarray}
\pi^{(k,l)}(t_0)&=&\frac12\pi^{(k)}(h)\otimes\1_{2l+1}\nonumber\\
\pi^{(k,l)}(t_+)&=&i\pi^{(k)}(e)\otimes\1_{2l+1}\nonumber\\
\pi^{(k,l)}(u_0)&=&\1_{2k+1}\otimes\frac12\pi^{(l)}(h)\nonumber\\
\pi^{(k,l)}(u_+)&=&\1_{2k+1}\otimes i\pi^{(l)}(f)\nonumber
\end{eqnarray}
and with respect to the direct product basis
$|k,l;m_k,m_l\rangle=|k,m_k\rangle\otimes|l,m_l\rangle$, where
$-k\le m_k\le k$ and $-l\le m_l\le l$, one obtains the $2l+1$ common
eigenvectors for $\sol_2(e)$, given by
\begin{equation}\label{pit}
  \pi^{(k,l)}(t_0)|k,l;k,m_l\rangle=k|k,l;k,m_l\rangle,\quad
  \pi^{(k,l)}(t_+)|k,l;k,m_l\rangle=0
\end{equation}
for $m_l=-l,-l+1,\ldots,l$. The $2k+1$ common eigenvectors for $\sol_2(f)$ are
given by
\begin{equation}\label{piu}
  \pi^{(k,l)}(u_0)|k,l;m_k,-l\rangle=l|k,l;m_k,-l\rangle,\quad
  \pi^{(k,l)}(u_+)|k,l;m_k,-l\rangle=0
\end{equation}
for $m_k=-k,-k+1,\ldots,k$. Thus, the $\sol_2$-invariant subspaces of the
representations of the proper Lorentz group, represented by the two
components of the Kronecker sum as ``left-handed'' and ``right-handed''
states, lead to the concept of helicity. Accordingly, the (group-theoretical
version of the) ``Weinberg ansatz'' is based on the concept of helicity.

\subsection{The Weinberg ansatz}\label{sec7}
Considering the states $(m_k,m_l)$ as points on a lattice of dimension
$(2k+1)\times(2l+1)$, according to equations~(\ref{pit}) and~(\ref{piu}), the
eigenstates of $\sol_2$ are found with the values $m_k=k$ and $m_l=-l$, i.e.\
at the boundary of this lattice. This can be interpreted physically as a
constraint on the spin degrees of freedom of the massless particle to only one
of the helicity states. It is even possible to show that for particles, this
is the left-handed helicity state~\cite{Groote:2022}. If, however, the
particle is equal to its antiparticle, the full spectrum of helicity states
is available. In general terms, this is formulated in
Refs.~\cite{Ohnuki:1976,Weinberg:1995} in the following way:
\begin{enumerate}
\item If a massless particle is equal to its antiparticle, it is described by
the irreducible representation $(k,k)$ of the proper orthochroneous Lorentz
group (Majorana case).
\item If a massless particle is not equal to its antiparticle, the particle
is described by the irreducible representation $(k,0)$ of the proper
orthochroneous Lorentz group, while the antiparticle is described by the
irreducible representation $(0,k)$ of the proper orthochroneous Lorentz group
(Dirac case).
\end{enumerate}
Note that the massless particle is defined via the Borel subgroup by the
irreducible representation of the proper orthochroneous Lorentz group without
the need to introduce space inversion.

\subsection{The Majorana case}
For the representation $(k,k)$, the ansatz yields $2k+1$ helicity states 
associated with $\sol_2(e)$,
\[|k,k;k,-k+p\rangle,\qquad p=0,1,\ldots,2k,\]
and $2k+1$ helicity states associated with $\sol_2(f)$,
\[|k,k;-k+p,-k\rangle,\qquad p=0,1,\ldots,2k.\]
Since the state $|k,k;k,-k\rangle$ is twice and at the same time excluded by
the condition
\[D_3^{(k,k)}|k,k;k,-k\rangle=0,\quad
  B_3^{(k,k)}|k,k;k,-k\rangle=2k|k,k;k,-k\rangle,\]
the particle with zero mass and helicity $\lambda=2k$ has $4k$ helicity
states. In particular, the defining representation $(\frac12,\frac12)$
describes a massless particle with helicity $1$ and two helicity states
\[|\Frac12,\Frac12;\Frac12,\Frac12\rangle\quad\text{and}\quad
  |\Frac12,\Frac12;-\Frac12,-\Frac12\rangle.\]

\subsection{The Dirac case}
According to Weinberg, the Dirac case $(k,0)\oplus(0,k)$ is a particular case
of the general situation $(k,l)$ with $l=0$ or $k=0$, respectively. For the
representation $(k,0)$ there exists only a single eigenvector
$|k,0;k,0\rangle$ of the Borel algebra $\bor_{1,3}(k,0)$, i.e.\ only a single
helicity state $|k,0;k,0\rangle$ with
\[t_0|k,0;k,0\rangle=k|k,0;k,0\rangle,\quad
t_+|k,0;k,0\rangle=u_0|k,0;k,0\rangle=u_+|k,0;k,0\rangle=0.\]
Similarly, the only helicity state for the representation $(0,k)$ is
$|0,k;0,-k\rangle$ with
\[u_0|0,k;0,-k\rangle=k|0,k;0,-k\rangle,\quad
  u_+|0,k;0,-k\rangle=t_0|0,k;0,-k\rangle=t_+|0,k;0,-k\rangle=0.\]

For example, the fundamental representation $\bor_{1,3}(\frac12,0)$ can be
expressed as
\[b_0(\Frac12,0)=\Frac12h,\quad b_1(\Frac12,0)=e,\quad
  b_2(\Frac12,0)=-ie\quad b_3(\Frac12,0)-\Frac12ih.\]
The corresponding representation of the algebra $\sol_2(e)$ has the form
\[t_0(\Frac12,0)=\Frac12h,\quad t_+(\Frac12,0)=ie\]
with $\sol_2(f)$ being trivial. Therefore, in the case of the irreducible
representation $(\frac12,0)$, there exists only a single solution
$e_1=(1,0)^T$, i.e.\ a helicity state $\lambda=1/2$, and this helicity state
is equal to the solution of the Weyl equation
\[\tilde\sigma_\mu p^\mu\psi(p)=0.\]
In case of the representation $(0,\frac12)$, $\sol_2(e)$ is trivial, and the
nontrivial algebra $\sol_2(f)$ is of the form
\[u_0(0,\Frac12)=-\Frac12h,\quad u_+(0,\Frac12)=-if\]
with only a single common eigenvector $e_1=(0,1)^T$.

\section{Conclusions}\label{sec8}
With this work, we have delved into the rich solvable structure of the proper
Lorentz group. As for a massless particle, the stabiliser subgroup of the
momentum four-vector is given by the Borel subgroup as the maximal noncompact
subgroup of the Lorentz group, of which the Lorentz group contains
two copies. Thus, we can generate the Borel subgroup as a Kronecker sum of two
copies of the simplest solvable algebra $\sol_2\subset\sL_2$ and,
correpondingly, the proper Lorentz group as a Kronecker sum of two copies of
the simplest noncompact algebra $\sL_2$. This is formulated in
Lemma~\ref{lemma}. From our investigation in this paper, we conclude that if
there is a particle state with pure helicity or spin, the mass of this
particle is zero and the stabiliser group is the Borel subgroup, fixing the
line of the light-like propagation. Therefore, at least for the
electromagnetic field, the symmetry determines the dynamics.

\end{document}